**On the influence of opacity variation on spatial structure of radiative shocks.**

M. Busquet,[a,b][*] M.Klapisch,[a] F.Thais [c]

[a] ARTEP,inc - Ellicott City, MD, USA

[b] Observatoire de Paris-Meudon - FRANCE

[c] CEA/DSM/IRAMIS/SPAM, 91191 Gif-sur-Yvette Cedex, FRANCE

**Abstract:**

We provide a theoretical analysis of Radiative Shocks, defined as supercritical shocks accompanied by an ionization wave in front of the density jump. In particular, we look at the influence of opacity variation with temperature and photon energy on spatial structure of radiative shocks, with a view to understanding a split precursor feature observed in recent experiments. We show that multigroup processing, a more refined angular description and improved low temperature opacities are needed to explore the radiative precursor structure, at least in some temperature regimes where rapid change of ionization can be found.

Corresponding author: Michel Busquet (busquet@this.nrl.navy.mil)

## 1. Introduction

Radiative Shocks (RS) [1] are strong, supercritical, shocks where the flux of ionizing photons coming from the shock front is large enough to launch an ionization wave in front of the shock. This radiative precursor (RP) is understood as a Marshak wave [2], which is a diffusive heat wave driven by photons in quasi-equilibrium with matter. The velocity of the RP wavefront increases with the ratio of the Rosseland mean free path to the material heat capacity at constant volume, $C_v$. Radiative Shocks are found in many astrophysical objects, exploding supernovae [3], supernova remnants [4], jets [5], and accretion shocks [6] driven by matter falling on the central star, funneled by the magnetic flux tubes. The spatial structure of the RP, its spectroscopic signatures, and the possibility of radiation front being unstable are not yet completely understood. These facts generate an interest in finding laboratory experiments that can emulate radiative precursors.

---
[*] Corresponding author. Email : busquet@artepinc.com

On earth, RS are found in high temperature blast waves in air and in the laboratories, in the so-called domain of "Laboratory Astrophysics", where the astrophysical phenomena are scaled to powerful ion and laser facilities. Radiating blast waves has been produced by the laser breakdown in residual gas inside the vacuum chamber [7]. Radiation driven ionization wave and RP have been produced with shock waves in foam [8] and in gas-filled small shock tubes at LULI [9], on OMEGA [10] and at PALS [11], where the shock is launched by a laser ablated foil acting as a piston.

The temperature rise in the RP, in front of the density jump in the purely hydrodynamic shock, leads to almost no increase of the bulk density. However, following the temperature- and radiation-induced ionization, the RP can be assessed by measuring the spatial profile of the electron density. Therefore, time-resolved visible interferometry is a convenient probing tool. Recently the first long duration interferometric observation, over 40 ns, has been achieved on the PALS facility [12], thanks to high optical quality targets [13]. In addition to the now usual detection of the temperature front, the build-up of an electron density jump, e.g., see Figs. 3b and 4b, has been observed in the middle of the RP. This is different from the apparition of a secondary shock launched in front of a decelerating RP [14].

Here we study the influence of the opacity variations for a range of temperatures or photon energies, that may explain this puzzling structure.

## 2. Experimental setup

The experiment we will study is the interferometric probing of a radiative shock created in a mini-shock tube by laser-ablated thin foils acting as a piston. The experiment setup, shown in Fig. 1, was performed at the PALS iodine laser facility in Prague, Czech Republic. We used the third harmonic of the laser at 438 nm, with a smoothing Phase Zone Plate (PZP) [15]. The PZP yields a flat-top focal spot of 0.55 mm FWHM in diameter, with 90% of the energy in a 0.7 mm spot, matching the inner section of the smallest shock tubes. The nominal pulse conditions are 0.35 ns duration and a 150-170 J energy at 438 nm after the infrared rejecting filter. Accounting for the reductions in transmission due to each of the PZP, the focusing lens, and the vacuum window the resulting energy on target is 110-130 J, which yields a mean laser intensity on target of $10^{14}$ W/cm$^2$. The targets [13], see Fig.2, are square tubes, 700 μm wide and 6 mm long, filled with xenon or a mixture of xenon and air, and closed by a gold coated 10 μm thin plastic foil. Laser ablation of the plastic foil accelerates the thin foil by the rocket effect, which launches a strong shock in the gas. The gold overcoat, facing the xenon, is used to stop the hot electrons and hard x-rays produced by the laser absorption minimizing the preheat of the shocked gas. Moreover, the gold foil contributes approximately half of the inertia of the target, resulting in a shock velocity that is insensitive to the temporal profile of the laser pulse.

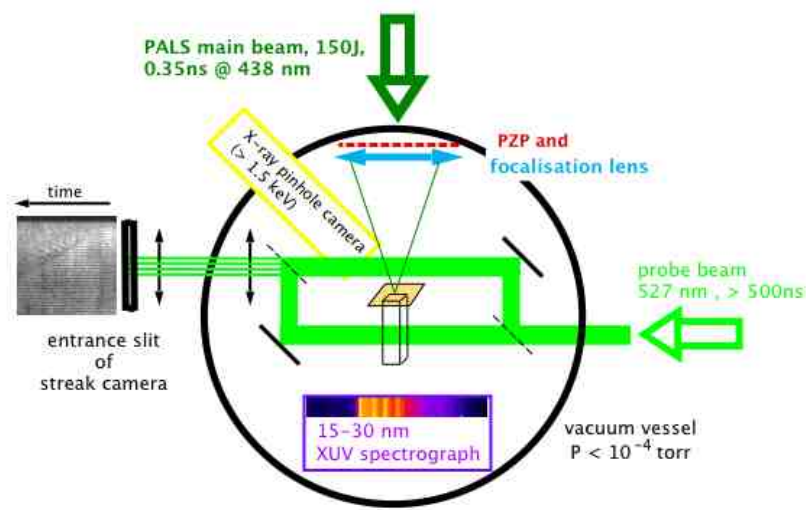

*Fig.1 : experimental setup (see text for details).*

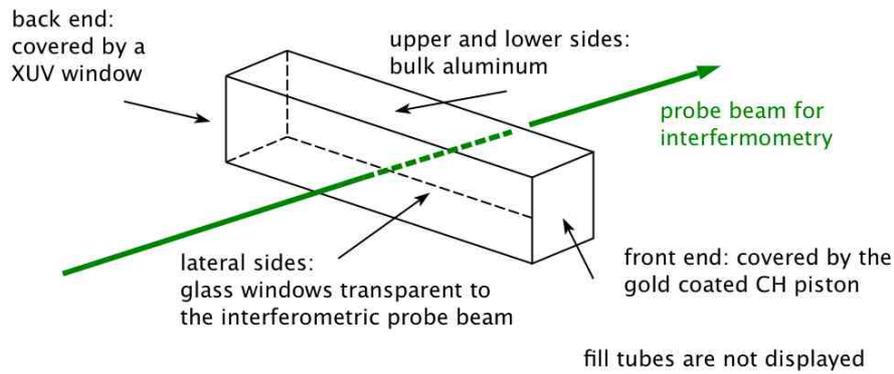

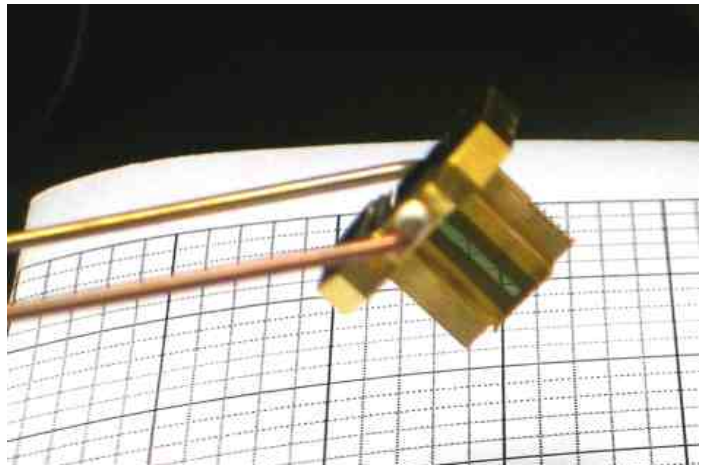

*Fig.2 : mini-shock tube used as targets, (upper) schematic , (lower) photograph.*

A Mach-Zehnder interferometer, set on a movable bench, was used to probe the electron density along the mini-shock tube. The probe beam has a 500ns duration pulse with a wavelength of $\lambda_P$ =527nm. The temporal evolution of the fringe pattern was recorded over a 100 ns sweep time by a streak camera with a field of view covering the full 6 mm length of the shock tube. In some shots, not presented here, the direction of observation was perpendicular to the tube allowing the observation of the transverse profile of the shock front [12]. We shall focus in this paper on two particular results, shown in Figs. 3 and 4, where a surprising feature, that we refer to as a split precursor (SP), identified by the thin black line in Fig. 3a, has been observed.

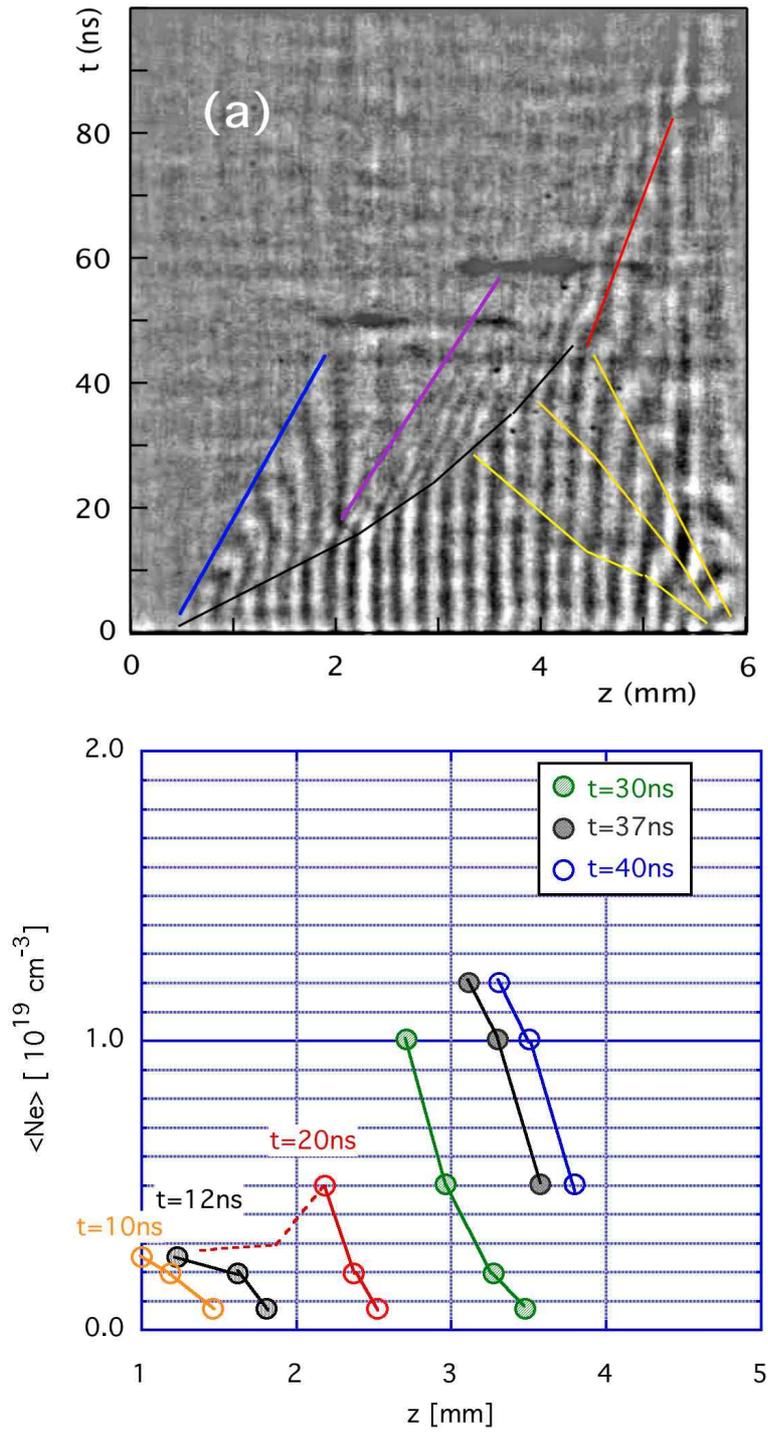

Fig.3 : contrast enhanced time dependant interferogram (upper) and deconvoluted electronic density profile (lower) for shot#1029_07 with a 0.1 bar Xenon fill . Lines in (a) display the limits of different waves, the precursor front is the thin black line.

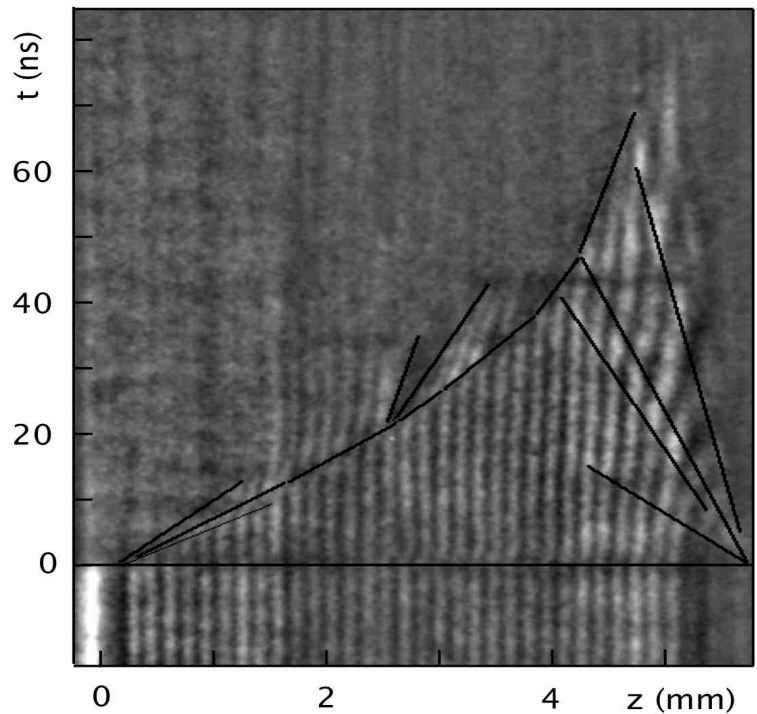

*Fig.4 : same as Fig.3 for shot#1030_05 with a 0.17 bar, 50% Xenon – 50% air, fill.  Horizontal and vertical axis ranges (resp. position and time) are identical to those in Fig.3a*

### 3. Deciphering interferograms

The main diagnostic of such experiments is the time-resolved interferograms shown in Figs. 3a and 4a. The shock tube is placed in one arm of a Mach-Zehnder interferometer, and imaged onto the entrance slit of a visible-light streak camera. In the figures, the horizontal axis is the distance "z" from the initial position of the piston, from left to right, covering the full 6 mm length of the shock tube. The vertical axis is the time, increasing from bottom to top. The time of the laser pulse is indicated on the images by the short "shimmy" of the fringes that is attributed to electromagnetic perturbation to the streak tube. The laser pulse thus coincides with the beginning of the sweep in Fig. 3a, and arrives 15 ns into the sweep in Fig. 4a.

As the reading of these interferograms is not straightforward we present it in details. The probe laser beam wavelength, $\lambda_P$, is 0.527 μm and has a duration of >500ns that allows the measurement of the phase shift  P. P is proportional to $n(\lambda) d$, where $n(\lambda)$ is the optical index of refraction of the ionized gas and *d* is the transverse dimension of the tube, versus the position in the tube at successive times "*t*". An adjusted tilt of the beam splitter gives an initial dephasing, with a linear dependence to the distance *z*. Let us call $D^{t=0}$ the initial interfringe distance.

In the regime of interest, the optical index of the medium is due to the plasma refractive index

$$n(\lambda) = [1 - N_e / N_{c,\lambda}]^{1/2} \approx 1 - 0.5 \, N_e / N_{c,\lambda}$$

where $N_e$ and $N_{c,\lambda}$ are the electron density and the critical density at the wavelength $\lambda$.

The phase shift is then given by

$$P(z,t) = P(z,0) + d \, <N_e(z,t)> / N_{c,\lambda}$$

where

$$\langle N_e(z,t) \rangle = \frac{1}{d} \int_0^d N_e(y,z,t) dy$$

is the electron density averaged over the transverse dimension *d,* which is 0.7mm.

Thanks to the initial dephasing, the phase shift *P(z,t)* can be measured by counting the number of fringes, or fractions of a fringe, passing at a given position *z*.

An electron density gradient, with a constant gradient scale length *L* gives regular fringes, i.e., equally-spaced, for which the interfringe distance is related to *L* and $D^{t=0}$. If the wavefront is moving with a constant velocity, like a RP detaching from its generating hydrodynamic shock, the whole fringe pattern drift with the same velocity producing a set of equidistant parallel tilted fringes. The precursor foot, PF, see Fig.5 where it is indicated by dashes, can be understood as the limit between the unperturbed fringes and the parallel tilted fringes, and is clearly identified in the interferograms. If the wave front slows down, the limit of unperturbed fringes forms a continuous curve that is the usual pattern observed for a radiative precursor foot with progressive attenuation due to lateral radiative power losses [16]. Our results show that same pattern (line labeled "P" in Fig.3a), which unambiguously reveals the existence of the radiative precursor.

If a second wave appears, either from a detaching ionization wave, as in the schematic set of electron density profiles displayed in Fig. 5, or from a density build-up within the RP, a second set of tilted fringes with a different slope will appear, with a more or less flat region in between. Such a split precursor has been observed in the interferograms recorded in our 2007 experiment - see Figs. 3-4. [12]

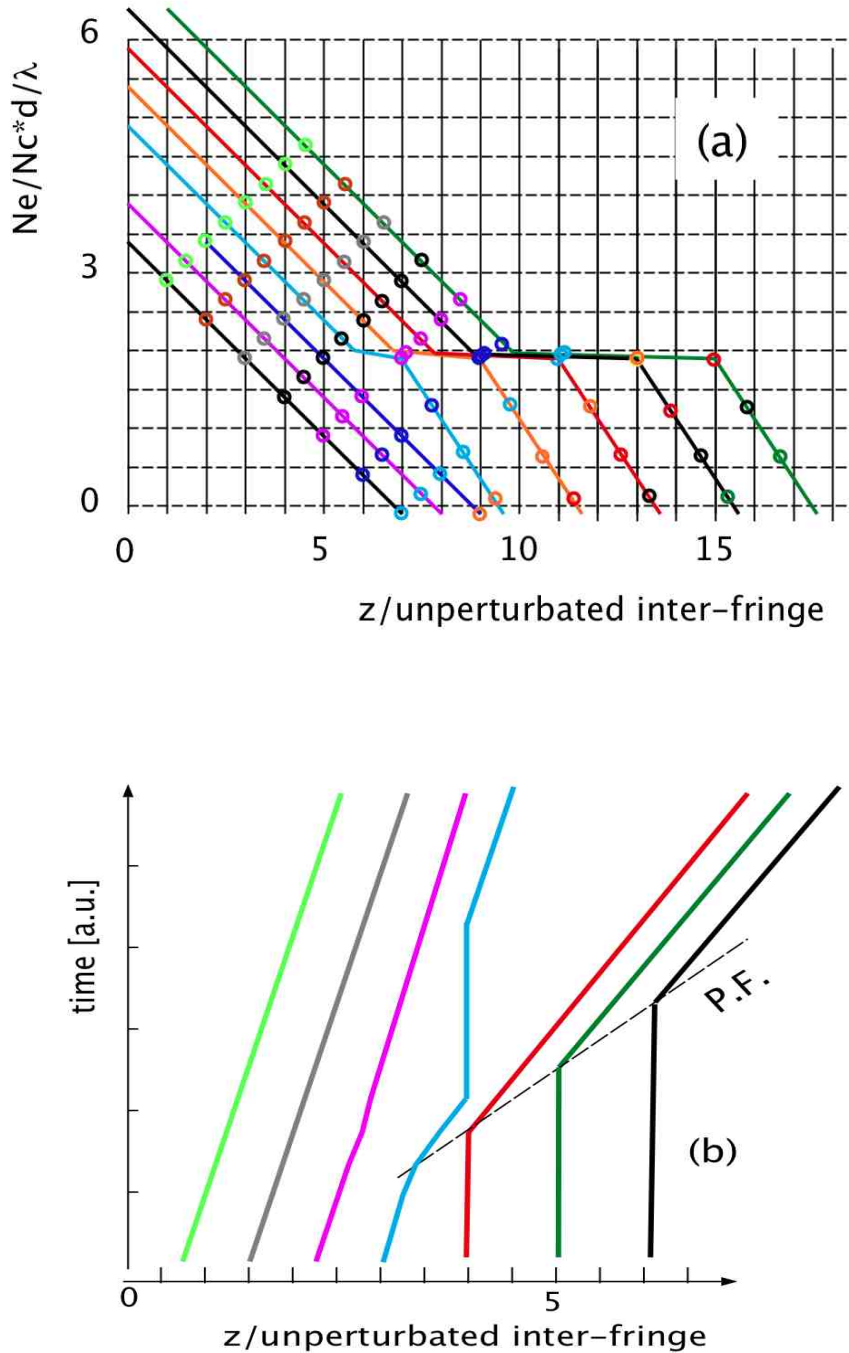

*Fig.5 fringes positions (b) derived from schematic density profiles (a) at successive times with an ionization wave detaching from the initial precursor. (color on line)*

In our simulations, we note also the existence of a reverse wave starting from the end of the shock tube, initiated by the heating up of the rear window, probably by the 438 nm light leaking through the piston, or from converted laser light in the wings of the PZP-smoothed focal spot going around the front window. Crossing of the direct and reverse radiative precursors impedes analysis after 45 ns.

With this introduction of the concept we can investigate two possible origins of the split precursor: 1) the variation of (average) opacities with temperature, or 2) the spectral description of the radiative transfer in the diffusion wave. We note that shock collapse, as described by Reighard et al [17], cannot explain the SP feature, as SP occurs in the precursor, well before the hydrodynamic density jump.

## 4. Synthetic interferograms: Varying the opacities

We present in this section results from hydrodynamic simulations including radiative transfer, with different description of the opacities. The simulations have been performed with the 1D rad-hydrocode EXMUL [19] derived from the MULTI code by Ramis et al. [18]. Radiation transport is computed by either the gray or the multigroup diffusion methods. The leading 3D effect is modeled by accounting for lateral radiative losses [20]. We set a wall albedo of 40% that is deduced from comparison of real 3D simulation using the HERACLES code [21] and experiments.[11] We start with a simulation of the whole target, including the laser absorption and the piston ablation. We then checked that the RP evolution computed with this full simulation is reproduced by a simulation using only the xenon gas but with imposed boundary velocities representing the piston movement, and a fixed aluminum wall at the end of the tube. Therefore, we present only the results from "imposed boundary velocities" simulations. We do not consider in this paper the possible preheating due to hard x-rays, which should be absorbed in the gold layer coating the back surface of the plastic foil. The results presented use the QEOS equation of state, which in a prior study was found to provide a reasonable xenon equation of state. [22]

For these simulations, opacities are needed for warm plasma, i.e., $T_e$ ranging from1 to 50 eV. This is a difficult temperature range as the number of relevant bound electrons configurations become extremely large. To overcome this difficulty, it is usual to switch to cold opacities, generally taken from Henke tables at the lowest temperature. Our new version STA-2010 [23] of the STA code of A. Bar-Shalom et al. [24] computes opacity at very low temperature, see Fig.6 for an example. STA-2010 uses a combination of the algorithm of Gilleron and Pain [25] and of our own algorithm [26] for the computation of partition functions that play an essential role in the STA model. This model includes a change to an average atom description with one superconfiguration at low temperature. It probably overestimates the average opacity below 4eV as the comparison with extrapolated values suggests and is supported by the results shown in Fig.7. On the other hand, Henke's cold opacities, even after addition of the free-free opacities, seem to strongly underestimate Rosseland means. Work in progress on STA2010 includes better description of pressure ionized atoms, with a gradual merging of free and bound states, and should overcome this problem. We compared three models, switching at low temperatures to (1) average atom, (2) extrapolation from higher temperatures, or (3) Henke + free-free, and looked at their effects on the RP structure. Note that at room temperature xenon is transparent for photon energies below 10 eV.

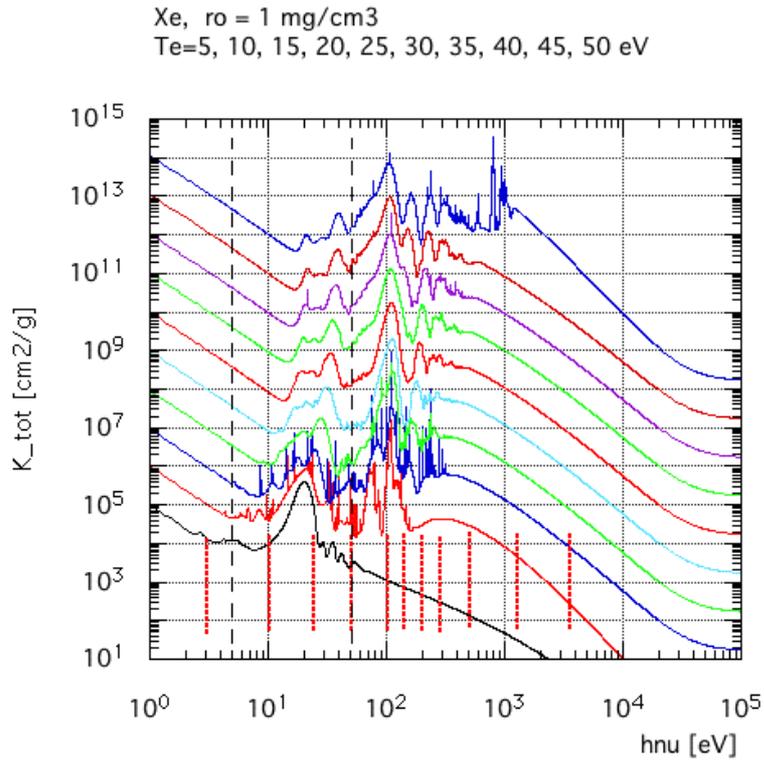

*Fig.6  Xenon spectral (or monochromatic) opacities computed with the STA-2010 code. Long vertical dashes show the temperature range relevant to our experiment. Short red dashes are the boundaries of a 13 groups frequency description.*

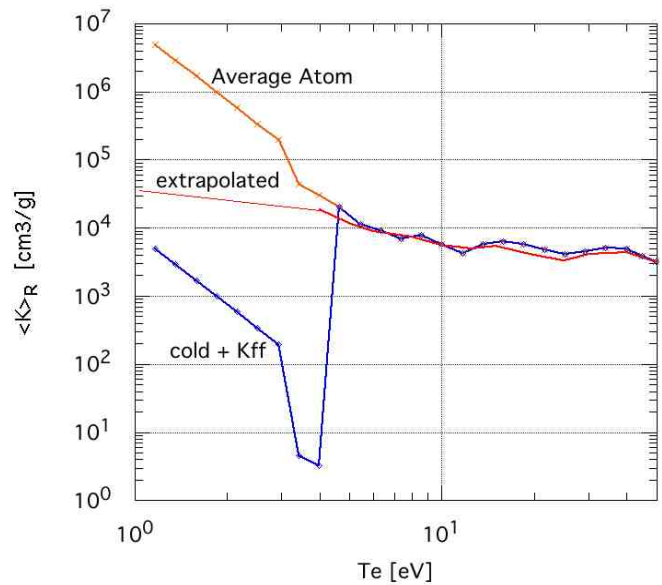

*Fig.7 : variation of the Rosseland mean with temperature for Xenon at a density of $1mg/cm^3$, with different hypothesis.*

First, we present in Fig. 8 computations performed with model 2 (extrapolated opacities at temperatures below 4eV) to illustrate effects of the multigroup description of the radiative transfer. The foot of the RP presents no more a sharp gradient when using multigroup description, as shown in Fig. 8b when compared to the Grey approximation, see Fig. 8a. "Suprathermal" photons yield a gentle rise of the temperature, the average ion charge <Z>, and the electron density. We also note a very rapid preheating, covering the full length of the tube. However, the P1 description of the multigroup model in EXMUL is a diffusion approximation –even for the high-energy groups responsible for this fast preheating. An angular description is needed, as in the M1 method used in the HERACLES code [21]. In Fig. 9, the opacity model "3" is used and gives a rapid variation of the mean opacity from the "cold-Henke" description at low temperature to a full description at higher temperature. The two temperature domains are sampled by the temperature gradient in the RP. Lower temperatures, with lower average opacities, give a large mean free path, an initially rapid rise and a large velocity (6mm/40ns). Higher temperature, with higher average opacities, give a small mean free path and a smaller velocity. The latter even slows down and is caught up by the hydrodynamic shock (the density jump) before re-acceleration at 50ns. A double RP is observed due to this sharp variation of average opacity with temperature. Both fronts are smoothed by the multigroup processing but are still presents as is show in Fig. 9b. However, in a higher temperature domain, when the RP temperature gradient covers less of the low opacity domain, this double front is no longer seen when using the gray radiative transfer, which is apparent in Fig. 10a, but remains apparent with multigroup radiative transfer as seen in Fig. 10b. Again the M1 radiative transfer appears to be required to correctly predict the RP structure in these cases.

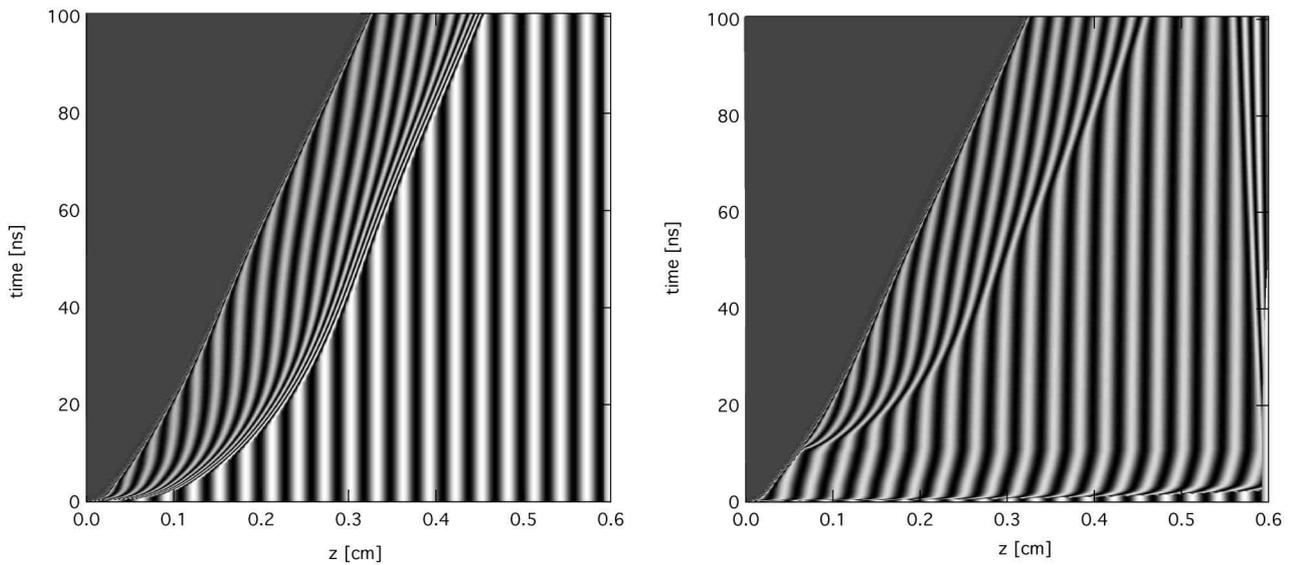

*Fig.8 synthetic interferograms obtained with model 2 of Xenon opacities in grey approximation(a) and multigroup processing (b)*

## 5. Conclusion

Our new version of STA is now able to compute average opacities down to temperatures of 1eV, although pressure ionization effects have to be supplemented by a progressive merging of free and bound states. Consequently, we are now able to study RP structure in the laboratory astrophysics framework. Although we did not determine the origin of the observed split precursor, we have shown that rapid variation with the temperature of average opacities yields a double RP structure. We have also shown that multigroup radiative transfer, an angular description and improved low temperature opacities are needed to explore the RP structure, when rapid change of ionization can be found.

## 6. Acknowledgments

This work was supported by University of Michigan, Ann Arbor, under the cooperative agreement No. DE-FC52-08NA28616, and by USDOE under a contract with Laser Plasma Branch, Naval Research Laboratory, Washington, DC. The experiment described here received financial supports from the